\documentclass[a4paper]{article}
\usepackage[affil-it]{au thblk}
\usepackage{hyperref}
\hypersetup{breaklinks=true,colorlinks,urlcolor=cyan,citecolor=green,linkcolor=blue,filecolor=black}

\title{On arXiv moderation system}
\author{Z.K. silagadze}
\affil{Budker Institute of Nuclear Physics SB RAS and 
Novosibirsk State University, 630 090, Novosibirsk, Russia.
silagadze@inp.nsk.su}

\date{}

\begin{document}

\maketitle
\begin{abstract}
The advent of arXiv has revolutionized scientific communication. However, its cultural significance goes far beyond simply accelerating scholarly communication.  The arXiv gave a powerful impetus to the democratization of science, freeing young scientists and not only, especially from totalitarian countries, from authoritarian oppression. Many of arXiv's innovative features have been blurred by the introduction of a moderation system. Without a doubt, a moderation system is essential to maintain the quality of arXiv content. However, I believe that it can be improved in line with arXiv's original intentions, using the very successful experience of the {\it MathOverflow} moderation system. 
\end{abstract}

\section{Introduction}
There are old anecdotes about Greek philosopher Thales that illustrate well the popular prejudices about science \cite{1,2,3}. According to the story told by Plato in his dialogue {\it Theatetetus} \cite{1,3}, Thales, studying the stars and looking up, fell into a pit and a Thracian servant-girl ridiculed him for helplessness in everyday affairs.

About two millennia later, the famous German philosopher Hegel gave a worthy answer to the one who mocked Thales as a person who has knowledge of celestial objects, but is not able to see what lies at his own feet \cite{4}:

``The people laugh at such things, and boast that philosophers cannot tell them about such matters; but they do not understand that philosophers laugh at them, for they do not fall into a ditch just because they lie in one for all time, and because they cannot see what exists above them."

In fact, anecdotes about Thales and Hegel's answer to one of them paint the popular image of scientists as eccentric but witty people living in their marble tower in the eternal sunshine of pure reason, revealing the truth by guesswork and refutation.  As is often the case, the reality is quite different \cite{5}:

``Science is not like that at all. There is a powerful establishment and a belief system. There are power seekers and career men, and if someone challenges the establishment he should not expect a sympathetic hearing.”

Also scientists are not like angels. Many prominent scientists had a difficult, if not unpleasant, character. An incomplete list includes such giants as Isaac Newton \cite{6,7}, William Shockley \cite{8}, John Nash \cite{9}, Lev Landau \cite{10}, Erwin Schr\"{o}dinger \cite{11,12}.

As Fyodor Dostoevsky aptly remarked: ``Yes, man is broad, too broad, indeed. I'd have him narrower. The devil only knows what to make of it!" \cite{13}. The great achievement of Western civilization is the realization that we should not rely on the prudence and good will of individuals (messiahs, kings, presidents, great scientists), but create social institutions that minimize the evil inclination inherent in man.

The same applies to the organization of scientific communication. The ideal organization should balance the meanness of spirit and madness that so often permeates human relationships, and maximize the noble features and sanity that, fortunately, are also inherent in people. 

\section{ArXiv and the freedom of academic expression}
The creation of arXiv in 1991 \cite{14,15,16} was without a doubt a revolutionary event. There were a lot of good things about arXiv when it first came out. Some of them, unfortunately, faded out over time. I briefly comment about two such features that I consider very important for academic freedom. 

When creating arXiv, the initial goal was to ``eliminate some of the inadvertent unfairness of the paper preprint distribution, where advance access to information was not uniformly available due to geography or institutional hierarchy" \cite{15}. What happened, however, was that arXiv instantly eliminated many other unfair practices (unintentional or not) in the production and dissemination of scientific information.

The current system of science management is not ideal, to put it mildly. ``The  system,  far  from  allowing  free  publication  of  results  among  professionals,  works  hand  in  hand  with  censorship.  Theoretically,  it  was  conceived  as  a  quality  control  but  its  functions  are  frequently  extended  to  the  control of power" \cite{19}. ArXiv has made the system much more democratic and almost free from the control of informal academic authorities. But democracy also has a downside. The need for quality control prompted arXiv to unfortunately introduce the endorsement system \cite{20}, a very serious step backwards from democracy, albeit a forced one since there was a danger that pseudo-scientific and near-scientific materials would flood arXiv.

Another big advantage of early arXiv was that ``Preprints speed up the distribution of research results and leave it to the reader’s discretion whether they are groundbreaking and of value" \cite{21}. Now this advantage is gone, as arXiv moderators, not readers, decide whether arXiv submissions are of value,  and this seriously undermines the original concept of arXiv \cite{22}. With such a  moderation policy, arXiv risks becoming a ``closed club" for what it considers to be ``mainstream" science \cite{23}. 

In addition, submissions can be put on hold due to questions about the correct classification, sometimes leading to significant delays in publishing the submission in arXiv, which is annoying.  The management of arXiv is well aware of such complaints, but they have to filter out submissions that do not meet certain standards in order to avoid arXiv's reputation as a place that attracts fringe science. The goal is ``to provide an open research sharing platform that balances speed and quality affordably" \cite{22} and it's fair to say that arXiv has done quite a good job of this so far. However, I believe that the scientific community would benefit if this imperfect system of moderation and endorsement could be improved and replaced with a more dynamic and efficient system that allows for more freedom without compromising the quality of the submissions.

\section{Advantages and disadvantages of peer review}
Without doubt, the peer review process was initiated in good faith to scrutinize and therefore improve academic papers submitted for publication. The moderation system used by arXiv is a light form of peer review and hence inherits all its advantages and disadvantages. 

Peer review is now widely used by scientific journals to ensure that journals publish high quality scientific information. While this fact indicates the overall effectiveness of the peer review process, the process has identified shortcomings and is considered a biased process \cite{27}.  The peer review process produces both excellent reviews and unhelpful or unprofessional reviews. Peer review is a social process influenced by the cultural background of the reviewers, and many of the influencing factors are still poorly understood \cite{27A,27B}.

Of course, the quality of peer review depends on the quality of the reviewers. And we have some outstanding examples that show how the peer review process should ideally work.
In 1936 Einstein submitted a paper to the Physical Review entitled ``Do Gravitational Waves Exist?" with Nathan Rosen as coauthor. The paper's conclusion was that gravitational waves do not exist, since the authors failed to find an exact solution to the Einstein equations corresponding to plane gravitational waves without introducing singularities into the metric.  Physical Review editor John Tate  sent the manuscript to a reviewer now known to be Howard Percy Robertson  \cite{28}. Thanks to his familiarity with the gravitational wave literature, which was more complete than Einstein's, Robertson found that Einstein's arguments were flawed and wrote a detailed account explaining the origin of singularities. Tate returned the manuscript to Einstein, along with a critical review, and politely requested that the various comments and criticisms of the reviewer be taken into account. Einstein angrily brushed off the reviewer's comments and resubmitted the manuscript to another journal, where it was immediately accepted in its original form. Fortunately, Einstein was saved from the embarrassment of publishing the wrong article by a chance meeting between Robertson and Einstein's new assistant, Leopold Infeld, at which Robertson explained the nature of the error and convinced Infeld of it. As a result, the article appeared with radically changed conclusions made by Einstein at the last moment (in proofs) in {\it Journal of the Franklin Institute} \cite{28}.

Of course, the Einstein-Robertson episode is the shining pinnacle of the peer review system, and we cannot expect every reviewer to be of Robertson's caliber (and not every author is comparable to Einstein). However, a disturbing feature of the huge number of modern referee reports is their complete uselessness.

Robertson, in one of his reports on Einstein's paper to Tate, remarked that the author ``is a man of good scientific standing, and it would seem to me that if he insists, he has more right to be heard than any single referee has to throttle!" \cite{28}. Of course, Einstein has earned the right to make a mistake, as he himself once remarked with a smile \cite{29}. But shouldn't lesser known scientists be allowed to make mistakes? I don't mean obvious mistakes. But subtle errors are a natural part of scientific research, especially in the field of innovation, and in my opinion, the scientific community should be more tolerant in this regard. In support of this thesis, we can recall the story of Joseph Weber's ``discovery" of gravitational waves \cite{30,31,32,33}.

Weber was a pioneer in the experimental study of gravitational waves. He started this whole field at a time when many prominent physicists did not even believe in the existence of gravitational waves. As his second wife, Virginia Trimble, aptly noted, ``Weber’s goal was to bring the Einstein equations into the lab. He felt, and I think it’s fair to say, he succeeded in doing that" \cite{34}.

After several years of intense, very thorough, and incredibly skillful experimental work, Weber claimed in 1969 that he had observed gravitational waves and became instantly famous. However, his fame soon faded to a large extent into notoriety when other experimental groups were unable to replicate his results. He did not admit that he was wrong and continued to insist until the end of his life that he saw exactly gravitational waves, despite clear indications that it was theoretically impossible to have such frequent and intense gravitational wave signals, as Weber claimed. Why was he so stubborn? Of course, he damaged his professional reputation with his position, but since no one has ever shown what exactly was wrong with Weber's experiment, we can say that Weber probably also behaved reasonably, continuing to argue his position \cite{30}.

Although it is now clear that Weber never observed gravitational waves, his experimental work was very important as it gave a huge boost to experimental and theoretical studies of gravitational waves, culminating in the observation of these waves by the LIGO collaboration in 2015 \cite{35}. Unfortunately for Weber and science, discussion of Weber's own experiments quickly dried up, leaving Weber on the fringes of the mainstream. This is certainly unfair and points to a flaw in the organization of science and its management.

That innovations can be difficult to understand, even for first-rate scientists, is well illustrated by the first perception of Feynman's path-integral approach to quantum theory and his now famous diagrammatic technique \cite{36,37,38,39}. Feynman left the meeting disappointed and depressed. Although, as Schwinger rightly noted years later, Feynman diagrams ``brought calculations to the masses", initially, for many years, physicists around the world, especially the older generation, experienced significant difficulties in understanding this innovation \cite{37,38}.

It is clear that we cannot, as a rule, expect a fair judgment from reviewers about a work if it contains radically new ideas. Of course, a certain conservatism is a healthy integral part of scientific culture. ``In part, Bohr's reaction to Feynman's ideas portrays the skepticism that should be used in approaching new and unproven theories, but the story also highlights the essential need for non-conformity if scientific progress is to be made" \cite{39}. Unfortunately, the reproduction of conformists is an inevitable, albeit unintended, result of modern education. In Feynman's case, the story had a happy ending. Feynman simply decided to publish his work because he knew that his approach gave correct answers in those cases in which Schwinger's independent calculations produced the same results by a quite different method \cite{36}.

Not all such stories have a happy ending, however. According to some sources, among them the famous American radiochemist Martin Kamen, one of the discoverers of carbon-14, who made a revolution in biochemistry, in 1921 before De Broglie and Schrödinger, the American physicist Arthur Constant Lunn (1877-1949) from Chicago University allegedly derived the de Broglie relation and the time-independent Schrödinger equation and applied the latter to the energy spectrum of the hydrogen atom, but the paper was rejected by a Physical Review referee (Gordon Fulcher) as too abstract and of no interest to Physical Review readers \cite{40,41,42,43,44}. The rejection was likely to have disastrous consequences for Lunn's scientific career as a physicist: ``Fulcher replaced Lunn as a member of Physical Review's editorial board early in 1922, and Lunn went on to withdraw bitterly from contact with most physicists" \cite{42}. Unfortunately, Lunn's manuscript has not been preserved. After Lunn's death, the mathematician and his co-author of many publications (including the very influential \cite{45}) J.~K.~Senior became responsible for Lunn's scientific legacy. He destroyed Lunn's quantum paper, stating that if he couldn't understand it, then no one else could \cite{43}. Without the manuscript at hand, it is impossible to judge how significant is the claim that Lunn anticipated Schr\"{o}dinger in wave mechanics. In any case, Lunn's story seems to indicate a serious failure of the moderation/peer review system.

\section{{\it MathOverflow}'s moderation system}
\href{https://mathoverflow.net/}{\it MathOverflow}  is a question and answer website used by professional mathematicians to discuss and gain mathematical knowledge \cite{46,47,48}. {\it MathOverflow}  was launched in 2009 by Anton Geraschenko, Scott Morrison and Ravi Vakil (and some Berkeley graduate students) \cite{49,50}.

{\it Mathoverflow}, while having a fairly rigid moderation system, is nevertheless very effective in providing quick informal communication and high-quality problem solving \cite{51}. The main points revealing the core philosophy behind the {\it mathoverflow} moderation system are described in \cite{52,53}.

In my opinion, the most important feature of the {\it Mathoverflow} moderation system is its dependence on the community. At the same time, the moderation process is open only to experienced users with a high reputation, with different levels of privileges depending on the reputation thresholds. For example, accessing moderator tools requires 10,000 reputation points \cite{54}.

Why are such changes important in the peer review system? As already mentioned, the moderation of scientific research is influenced by the historical era with its emerging communication technologies. For example, the first scientific journal was created in 1665 by Henry Oldenburg as a result of the appearance of the printing press. The system of peer review and editorial review emerged simultaneously as a filter to ensure the high quality of research in response to an increase in the volume of scientific correspondence \cite{55}. 

We are now witnessing a transition from paper-based communication among scientists to an Internet-based world. We cannot expect the old peer review system that developed in the paper world to remain adequate in the new Web era. It may well be expected that in this new era, ``the reviewer will metamorphose from gatekeeper to interlocutor and collaborator" \cite{55}. Therefore, I think the future arXiv moderation system should emphasize self-organization even more than it takes place in {\it Mathoverflow}.

It can be feared that excessive self-organization can lead to chaotic behavior. The following interesting example shows that such concerns are groundless.

The bus system of the city of Cuernavaca in Mexico, unlike other metropolitan bus systems, is completely decentralized without any timetables (at least it was when the two Czech physicists Milan Krb\'{a}lek and Petr \v{S}eba studied it in 2000 \cite{56}). Each bus is the property of the driver and they try to maximize their income by maximizing the number of passengers they carry. At each bus stop, drivers receive, for a fee, information from involved bystanders about the departure time of the previous bus, and using this information, drivers self-adjust their speed to optimize the distance to the previous bus. It is noteworthy that such a simple optimization process produces almost the greatest possible social benefit for both bus drivers and passengers (see \cite{57}).

\section{Concluding remarks}
There are many sins in modern Western civilization (and Russian civilization, despite the current insane attempts to tear it away from the West, is its offspring and an inseparable part). However, at the same time, Western civilization is fraught with many valuable achievements of mankind, which have no analogues in history. Modern science and rationality are gems of Western culture that we must cherish at all costs. The important historical fact is that science and rationality flourished in societies that practiced greater tolerance in most areas of public life, and withered and perverted in societies where intolerance and religious fanaticism prevailed.

``It is important to keep in mind that the decline of scientific activity is the rule, not the exception, of civilizations. While it is commonplace to assume that the scientific revolution and the progress of technology were inevitable, in fact the West is the single sustained success story out of many civilizations with periods of scientific flourishing. Like the Muslims, the ancient Chinese and Indian civilizations, both of which were at one time far more advanced than the West, did not produce the scientific revolution" \cite{60}.

The improvement of the arXiv moderation system based on self-organization will become a small brick in the sustainability of the monumental building of a rational point of view, which is the basis of a more just and reasonable world. That's why I urge arXiv administrators to pay attention to the very successful {\it Mathoverflow} moderation principles. In particular, the following improvements to the arXiv moderation system can be proposed:
\begin{itemize}
\item Anyone can submit a manuscript, as it was when arXiv first appeared. Potential challenges of such openness will be compensated by other features of the moderation system.
\item The ArXiv community is organized around a reputation system that can be earned through the quality of submissions rated by other users.
\item Only users with a high enough reputation can participate in moderation with varying privilege levels based on reputation thresholds.
\item The arXiv leadership openly and democratically selects a number of core moderators who lead the moderation process.
\item The principal moderators do  as  little  as  possible,  but  when  necessary,  their actions are powerful and highly concentrated. In particular, any violation of academic ethics is immediately and severely punished.
\item A particular submission is removed from the repository when a certain threshold of votes for removal is reached, but never based on the opinion of only one user, even the principal moderator.
\item When a certain threshold of removed submissions from a particular user is reached, the user is automatically blocked for a set period of time, which increases twice if the malicious behavior is repeated after the block period ends.
\end{itemize}

\section*{Acknowledgements}
I first heard the story of Arthur Lunn's quantum paper from Peter Horvathy (who learned about it from Gary Gibbons). I thank him for reading the manuscript and for his support. Helpful comments by Jalel Alizzi helped improve the manuscript. I also thank to Mikhail Shifman for critical remarks. The first version of the manuscript \cite{61} was more like an essay than an academic article. I thank anonymous reviewers whose comments helped to focus the manuscript on its most significant aspects.

\end{document}